\documentclass[english,aps]{revtex4}
\usepackage[T1]{fontenc}
\usepackage[latin1]{inputenc}
\usepackage{amsmath}
\usepackage{graphicx}
\usepackage{amssymb}
\usepackage{esint}

\makeatletter
\@ifundefined{textcolor}{}
{%
 \definecolor{BLACK}{gray}{0}
 \definecolor{WHITE}{gray}{1}
 \definecolor{RED}{rgb}{1,0,0}
 \definecolor{GREEN}{rgb}{0,1,0}
 \definecolor{BLUE}{rgb}{0,0,1}
 \definecolor{CYAN}{cmyk}{1,0,0,0}
 \definecolor{MAGENTA}{cmyk}{0,1,0,0}
 \definecolor{YELLOW}{cmyk}{0,0,1,0}
 }

\makeatother

\makeatother

\usepackage{babel}

\begin{document}

\title{\textsf{Analytic behavior of the QED polarizability function at finite
temperature}}

\author{A. Bernal}

\affiliation{Dept. de Matemàtica Aplicada i Anàlisi, Universitat de Barcelona.
\\
Facultat de Farmàcia Av Joan XXIII s/n Edifici A, Escala A, Tercer
pis, Matemàtiques 08028, Barcelona, Spain}

\author{A. Pérez}

\affiliation{Departament de Física Teòrica and IFIC, Universitat de València-CSIC
\\
 Dr. Moliner 50, 46100-Burjassot, Spain\\
 }
\begin{abstract}
We revisit the analytical properties of the static quasi-photon polarizability
function for an electron gas at finite temperature, in connection
with the existence of Friedel oscillations in the potential created
by an impurity. In contrast with the zero temperature case, where
the polarizability is an analytical function, except for the two branch
cuts which are responsible for Friedel oscillations, at finite temperature
the corresponding function is not analytical, in spite of becoming
continuous everywhere on the complex plane. This effect produces,
as a result, the survival of the oscillatory behavior of the potential.
We calculate the potential at large distances, and relate the calculation
to the non-analytical properties of the polarizability. 
\end{abstract}
\maketitle

\section{Introduction}

The potential created by a static ionic impurity in metallic alloys
has been considered by many authors (see, for example \cite{FetterWal,Mahan}).
At large distances, the potential shows an oscillatory behavior, damped
as negative powers of the distance $r$, a phenomenon which is known
as Friedel oscillations \cite{FR52}. At zero temperature, this kind
of behavior has been associated to the existence of the Kohn singularity
\cite{KO59} in the quasi-photon polarizability, induced by the sharp
edge of the degenerate electron distribution. At the threshold of
the interaction, i.e. at the threshold of electron-hole creation,
the momentum of the quasi-photon is equal to $2p_{f},$ the diameter
of the Fermi sphere, with $p_{f}$ the Fermi momentum of the electrons.
It must be noticed that the quasi-photon self-energy presents a singularity
at this point. The presence of a singularity allows to obtain, with
the help of Lighthill's method, the asymptotic form ($r\rightarrow\infty$)
of the potential, as an expansion in terms of the form $\cos(2p_{f}r)$
and $\sin(2p_{f}r)$, damped as negative powers of $r$, and enhanced
by powers of $\log(4p_{f}r)$ \cite{LI64}.

Alternatively, one can perform an analytical continuation of the quasi-photon
polarizability to complex values of the transferred momentum $q$,
and do the integration by deforming the circuit in the complex plane.
Aside from the {}``Debye pole'' of the quasi-photon propagator on
the imaginary axis, giving an exponentially damped contribution, the
polarizability has two branch cuts starting at $q=\pm2p_{f}$, which
are responsible for the long-distance oscillatory behavior \cite{FetterWal,KA88}
. In other type of plasmas, one can also find additional complex poles
in the propagator, giving raise to an exponentially damped oscillatory
contribution \cite{SPD01}.

Let us now consider the situation at finite temperature $T$. Since
the electron distribution is spread out, from energy and momentum
conservation in the collisions between the electrons and soft quasi-photons
we expect that screening becomes more effective than in the zero-temperature
case. This is, in fact, the case and one obtains that the oscillations
are still present, although they are damped as temperature increases
\cite{FetterWal}. Of course, this is a desirable result, showing
that the limit $T=0$ is not pathological.

We would like to understand this aspect from a mathematical point
of view. On one side, the above discussion, based on the Kohn singularity,
does not hold, since we know that the singularity disappears at $T\neq0$
\cite{SPD01}. On the other side, one can be tempted to extend the
integration to the complex $q$ -plane. If the $T>0$ polarizability
is still an analytical function, one has to face these possibilities:
either the function has discontinuities on the complex plane (i.e.
branch cuts, as in the $T=0$ case) or poles , or both. As we discuss
later, the first possibility does not appear: the polarizability becomes
a continuous function at non-zero temperature. On the other hand,
the appearance of additional poles, at an arbitrary small temperature,
which were not present at $T=0,$ would indicate a strong anomaly,
as branch cuts should be suddenly transformed into complex poles,
a possibility which looks too exotic and would point towards a singular
behavior with temperature. As discussed above, this is not the case.
Then, how can we account for the oscillations?

As we show in this paper, the explanation lies on the fact that, at
finite temperature, the polarizability is non-analytical over the
whole complex plane. In fact, this function can be understood as a
superposition of a family of functions, each one having discontinuities
at $q=\pm2p$. Each one of the functions on this family gives an oscillatory
result and the sum, for not too large temperatures, is still oscillatory.
As temperature increases, the range of values of $p$ that effectively
contribute is enlarged, leading to a destructive interference. For
this reason, oscillations are damped with temperature.

This paper is organized as follows. The analytical properties of the
polarizability function, both at $T=0$ and $T>0$, are discussed
in sections 2 and 3. In section 4 an expression for the potential
is obtained. Section 5 contains a brief discussion and concluding
remarks.

\section{The polarizability in the complex plane}

The potential created by a static ionic impurity in an electron gas
can be calculated from:

\begin{eqnarray}
V(r) & = & \frac{e^{2}}{\pi r}I\left(r\right)\nonumber \\
I\left(r\right) & = & \int_{-\infty}^{+\infty}dq\,\frac{q\sin(qr)}{\nu(q)}=Im\int_{-\infty}^{+\infty}dq\,\frac{q\exp(iqr)}{\nu(q)},\label{I(r)}\end{eqnarray}
 where %
\footnote{We use the system of units $\hbar=c=1$.%
}

\begin{equation}
\nu(q)=q^{2}-\lambda\chi^{T}(q),\label{nu(q)}\end{equation}
 $\lambda=4\pi e^{2}$, $e$ is the electron charge, and $\chi^{T}(q)$
is the polarizability at temperature $T$ within the RPA (random phase
approximation), which can be written as:

\begin{equation}
\chi^{T}(q)=\int_{0}^{\infty}dp\, g(p)\chi^{0}(q,p).\label{chiT}\end{equation}

Here, \begin{equation}
\chi^{0}(q,p)=m\frac{-4pq+\left(q^{2}-4p^{2}\right)\log|\frac{q+2p}{q-2p}|}{8\pi^{2}q}\label{realchi0}\end{equation}
 is the $T=0$ polarizability, $m$ is the electron mass, and \begin{equation}
g(p)=-\frac{\partial f(p)}{\partial p}\end{equation}
 where $f$ is the Fermi-Dirac distribution for the electrons \begin{equation}
f(p)=\frac{1}{1+\exp\left(\frac{\varepsilon\left(p\right)-\mu}{T}\right)}.\end{equation}

We consider a non-relativistic plasma. Therefore, in the latter formula
$\varepsilon\left(p\right)=\frac{p^{2}}{2m}$. Finally, $\mu$ is
the chemical potential. Under these conditions, we have: \begin{equation}
g(p)=\frac{p}{4mT}\frac{1}{\left[\cosh\left(\frac{\varepsilon\left(p\right)-\mu}{2T}\right)\right]^{2}}\label{functiong}\end{equation}

As mentioned above, Eq. (\ref{nu(q)}) corresponds to the RPA level.
A further improvement of this approximation can be introduced via
a local field correction. For zero temperature, simple analytical
modelizations are given e.g. in \cite{ICH82}. However, to our knowledge
no such analytical models exist at finite $T$. For this reason, since
we only intend to give a qualitative explanation for oscillations
at $T>0$, we stay at the RPA level. Of course, in order to obtain
more accurate results, one would need to go beyond this approximation.

The usual way to calculate the integral in Eq. (\ref{I(r)}) is by
performing an analytical continuation to the upper complex half-plane:

\begin{equation}
I(r)=Im\int_{C}dq\,\frac{q\exp(iqr)}{\nu(q)},\label{V(r)inC}\end{equation}
 where $C$ is now a circuit contained on the upper half-plane. In
doing so, we need to investigate the analytical properties of $\chi^{T}(q)$
(see Eq. \ref{nu(q)}) in the interior of $C.$ These properties are
defined by Eq. (\ref{chiT}).

\section{Analytical properties of the polarizability function}

We start by defining an extension to complex values of $q$ of the
function $\chi^{0}$ defined in Eq. (\ref{realchi0}). Let's define
$\chi^{0}(q,p)$ for $p>0$ and complex $q=x+iy$ by the following
formula:

\begin{equation}
\chi^{0}(q,p)=-\frac{mp}{2\pi^{2}}+\frac{m}{8\pi^{2}}\frac{(q-2p)(q+2p)}{q}\left\{ \log\frac{|q+2p|}{|q-2p|}+i\left[\arctan\left(\frac{y}{x+2p}\right)-\arctan\left(\frac{y}{x-2p}\right)\right]\right\} .\label{complexchi0}\end{equation}

We first determine a maximal domain in the complex plane where the
complex function $\chi^{0}$ is defined. For $p>0$ we define the
complex domains $G_{1,p}$, $G_{2,p}$ and $G_{3,p}$ to be respectively
the set of all complex $q=x+iy$ such that $\Re q<-2p$, $-2p<\Re q<2p$
and $\Re q<2p$, respectively . One can check that the expression
(\ref{complexchi0}) defining $\chi^{0}(q,p)$ converges to $-\frac{mp}{2\pi^{2}}$
as $q\to0$. We also consider $L$, the principal determination of
the complex logarithm \[
L(w)=\log|w|+i\arg w,\]
 where $\arg$ is the principal determination to the argument function,
taking values between $-\pi$ and $+\pi$. If $q$ is in $G_{1,p}$
we have the following representation: \[
\chi^{0}(q,p)=-\frac{mp}{2\pi^{2}}+\frac{m}{8\pi^{2}}\frac{(q-2p)(q+2p)}{q}\left\{ L\left[-\left(q+2p\right)\right]-L\left[-\left(q-2p\right)\right]\right\} .\]
 We also have similar representations for $q\neq0$ in $G_{2,p}$
and in $G_{3,p}$. It follows that $\chi^{0}(q,p)$ is analytic in
the domains $G_{p,i}$, excluding perhaps $q=0$ in $G_{p,2}$. Since
$\chi^{0}(q,p)$ converges as $q\to0$, it follows that $\chi^{0}(q,p)$
is analytic on the three domains $G_{p,i}$.

Thus $\chi^{0}(q,p)$ is analytic in the whole complex plane, excluding
the two branch cuts $x=\pm2p$. One can also check that $\chi^{0}$
can be continuously extended to the real points $q=\pm2p$. It is
also true that $\chi^{0}(q,p)$ can be continuously extended to any
of the closures $\bar{G}_{p,i}$, but those extensions don't match
up at the boundaries (except at the real points $q=\pm2p$) since
$\chi^{0}$ has jump discontinuities there.

We now consider $\chi^{T}$ defined as in (\ref{chiT}) using the
complex $\chi^{0}$ instead. In order to prove that the integral defining
$\chi^{T}$ exists and to study its analytical properties, (continuity
or computation of complex integrals) it is necessary to use the classical
theorems of Fubini and Lebesgue from integration theory. The key properties
of the function $g(p)$, (\ref{functiong}), that will be needed are
that $g$ converges to zero exponentially as $p\to\infty$ and that
$g(p)=O(p)$ as $p\to0$. These properties will allow us to dominate
products of $g(p)$ and not too fast growing functions of $p$. In
particular, if a function $h(p)$ has polynomial growth as $p\to\infty$
and is continuous, or has a behavior like $1/p$ or like $\log p$
as $p\to0$ , then the product $g(p)h(p)$ will be Lebesgue-integrable
and the necessary computations will make sense.

We first consider for which points $q=x+iy$ of the complex plane
does the integral in Eq. (\ref{chiT}) exist. First of all, since
$\chi^{0}(0,p)=-mp/\pi^{2}$, we have that $\chi^{T}(0)$ is well
defined by Eq. (\ref{chiT}). If we fix a point $q_{0}\neq0$, formula
(\ref{complexchi0}) can be used and we can bound $\chi^{0}(q_{0},p)$
by a polynomial in $p$. In this way, we see that the function $\chi^{T}$
is defined on the whole complex plane.

Turning to the continuity of $\chi^{T}$, we can start with a point
$q_{0}\neq0$ and a sequence $0\neq q_{n}\to q_{0}$. We consider
the compact set $K=\{q_{0}\}\cup\{q_{n},n=1,2,\ldots\}$ so that $0\notin K$.
We can also use Eq. (\ref{complexchi0}) to obtain a polynomial bound
to $|\chi(q,p)|$ uniformly on $q\in K$, so that Lebesgue's dominated
convergence theorem can be applied and the continuity of $\chi^{T}$
at $q_{0}\neq0$ can be established.

To prove the continuity of $\chi^{T}$ at $q=0$, we bound $\chi^{0}(q,p)$
in a slightly different way. We first consider $q=x+iy$ with $|q|<1$.
If $q\in G_{2,p}\setminus\{0\}$, we have \[
\chi^{0}(q,p)=-\frac{mp}{2\pi^{2}}+\frac{m}{8\pi^{2}}(q-2p)(q+2p)\psi_{p}(q,)\]
 where \[
\psi_{p}(q)=\frac{L(2p+q)-L(2p-q)}{q},\]
 $L$ is the principal determination of the complex logarithm as before,
and the function $\psi_{p}$ can be extended to an analytic function
in the whole domain $G_{2,p}$. One then checks that \[
\psi_{p}(q)=2L\left(1+\frac{q}{2p}\right).\]
 We can check the above equality by expanding both functions in power
series for small $\frac{q}{2p}$ and applying analytic continuation
to $G_{2,p}$. Then we have the estimate \[
\left|L\left(1+\frac{q}{2p}\right)\right|\le|\log|2p+q||+|\log2p|+\frac{\pi}{2},\]
 which allow us to bound $|\chi^{0}(q,p)|$ by the sum of a polynomial
on $p$ and a term which is the product of a polynomial on $p$ and
$|\log2p|$.

On the other hand, if $q\in G_{1,p}\cup G_{2,p}$, the estimate of
$\chi^{0}(q,p)$ as in (\ref{complexchi0}) gives the sum of two terms.
The first term involves the logarithmic part of (\ref{complexchi0})
and gives, after some inequalities, a polynomial bound. The second
term involves the {}``arctan'' part and, using the fact that $q\in G_{1,p}\cup G_{3,p}$
gives a bound by a polynomial in $p$ multiplied by $1/p$.

In any case, if $|q|<1$, $\chi^{0}(q,p)$ is bounded by terms that
are polynomial on $p$, logarithmic or $1/p$, uniformly on $q$.
Therefore, the product $g(p)\chi^{0}(q,p)$ is dominated by a Lebesgue-integrable
function and $\chi^{T}$ is defined and continuous in the whole complex
plane.

\begin{figure}
\includegraphics[width=8cm]{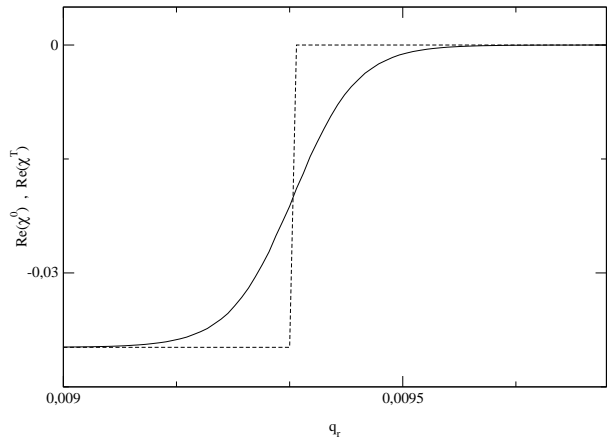}

\caption{The electron gas polarizability at zero temperature (dashed line),
for a value $q_{i}=1$, as a function of $q_{r}$. All magnitudes
are in units of the electron mass. For the chosen parameters (see
the text), we have $2p=0.00934$ . Also shown for comparison is the
finite-temperature polarizability (solid line), when $T=0.01T_{F}$.}

\label{dibujachi} 
\end{figure}

Thus, at finite temperature the above mentioned discontinuities disappear,
and the function $\chi^{T}(q)$ becomes continuous everywhere. This
is clearly illustrated in Fig. \ref{dibujachi}, where we have plotted
the real part of $\chi^{0}(q,p)$ (dashed line). We have chosen $p$
as the Fermi momentum of an electron gas with $r_{s}=3$ , where $r_{s}$
is the mean interelectronic distance in units of the Bohr radius,
as usual. We also plot (solid line) the real part of $\chi^{T}(q)$
for the same density, and a temperature $T=0.01T_{F}$ (here, $T_{F}$
is the Fermi temperature). As is apparent from this figure, the discontinuity
of $\chi^{0}(q,p)$ around $\Re(q)=2p$ disappears when the temperature
is non zero.%

It turns out that $\chi^{T}$ is non analytic at any point of the
complex plane. To show this, we can compute the integral

\begin{equation}
\int_{R}dq\,\chi^{T}(q)=\int_{R}dq\,\int_{0}^{+\infty}dp\, g(p)\chi^{0}(q,p),\label{integralchiT}\end{equation}
 where $R$ is a rectangle. If the rectangle is contained in the first
quadrant, standard estimates on $|\chi^{0}(q,p)|$ show that the double
integral of $\chi^{0}(q,p)$ is well defined, the needed change of
the order of integration is possible, and the usual procedures in
complex analysis show that the value of (\ref{integralchiT}) is not
zero. The computation can be reproduced in all the other quadrants
and we finally arrive to the conclusion that $\chi^{T}$ is not analytic
at any point.

We give the details of the computation in the first quadrant as an
illustrative example. Suppose that the rectangle $R$ is defined by
the vertices $q=x_{0}+iy_{0}$ and $q=x_{1}+iy_{1}$.

We introduce the notations \begin{eqnarray}
\chi^{0}(C_{2p}^{+}) & \equiv & \lim_{\epsilon\rightarrow0}\chi^{0}(2p-\epsilon+iq_{i},p),\nonumber \\
\chi^{0}(C_{2p}^{-}) & \equiv & \lim_{\epsilon\rightarrow0}\chi^{0}(2p+\epsilon+iq_{i},p),\label{discontin}\end{eqnarray}
 and the definition \begin{equation}
\Delta\chi^{0}(y,p)\equiv\chi^{0}(C_{2p}^{+})-\chi^{0}(C_{2p}^{-})=-\frac{m}{8\pi}\frac{y\left(y-i4p\right)}{\left(y-i2p\right)}.\end{equation}
 By deforming the contour integral, we can write \begin{equation}
\int_{C}dq\,\chi^{0}(q,p)=-i\int_{y_{0}}^{y_{1}}dy\,\Delta\chi^{0}(y,p).\end{equation}
 Therefore \begin{equation}
\int_{R}dq\,\chi^{T}(q)=-i\int_{x_{0}/2}^{x_{1}/2}dp\, g(p)\int_{y_{0}}^{y_{1}}dy\,\Delta\chi^{0}(y,p),\end{equation}
 which is different from zero.

\section{Computation of the potential}

We now return to the computation of $V(r)$, as given by Eq. (\ref{V(r)inC}).

\begin{figure}
\includegraphics[width=8cm]{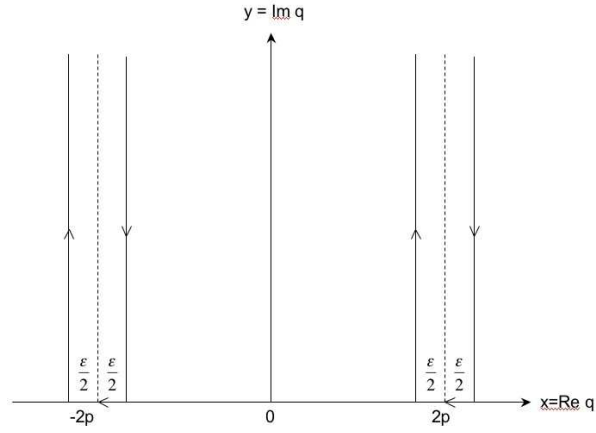}

\caption{Analytically-continued structure of the zero-temperature polarizability.
Represented schematically is the contour path used for integration.}

\label{circuitforV} 
\end{figure}

First, let us calculate \begin{equation}
\int_{C}dq\, q\exp(iqr)\chi^{T}(q)=\int_{0}^{\infty}dp\, g(p)\int_{C}dq\, q\exp(iqr)\chi^{0}(q,p),\label{interchange}\end{equation}
 where $C$ is the contour in figure \ref{circuitforV}.

In order to see that the above integral is well defined, to be able
to interchange the order of integration and to compute the limit when
the width of the contours in figure \ref{circuitforV} tends to zero,
it is necessary to estimate $\chi^{T}(q)$, and take into account
that the imaginary part of $q$ in the above contours is unbounded.

We can keep the width of the contours in figure \ref{circuitforV}
equal to $\epsilon\le\epsilon_{0}$. Then we can make the estimates
for the full rectangles of width $\epsilon_{0}$. It can be checked
that the expression $q\exp(iqr)\chi^{0}(q,p)$ admits a polynomial
bound in $p$ uniformly on $q$ inside those rectangles if $y\le|x|$,
and that it can be bounded by the product of a polynomial on $p$
by the exponential $\exp(-|q|r/2)$ if $y>|x|$. With those estimates,
it can be seen that, defining \[
I(p)=\int_{C}d|q|\,\left|qe^{iqr}g(p)\chi^{0}(q,p)\right|,\]
 it follows that $\int_{0}^{+\infty}\d p\, I(p)<+\infty$, and the
order of integration can be interchanged. Indeed, the expression \[
g(p)\left|\int_{C}dq\, qe^{iqr}\chi^{0}(q,p)\right|\]
 can be dominated with a Lebesgue-integrable function of $p$, so
that the limit when $\epsilon\to0$ can be taken inside the integral.

We now compute the value of (\ref{interchange}). In the following,
we leave aside the contribution from the Debye pole, which lies on
the imaginary axis. At zero temperature, its position is given by
$q_{D}=\pm i(p_{f}m\lambda)^{1/2}/\pi$. The pole contribution can
be easily incorporated in our calculations via the residue theorem,
giving rise to an exponentially damped term, which does not appreciably
modify the results for the long range behavior we are interested here.
Moreover, at finite temperature the pole is located at higher positions
on the imaginary axis \cite{Ya89,Sch02}, therefore giving an even
smaller contribution. %

Under this approximation, we can write \begin{eqnarray}
 &  & \int_{C}dq\, q\exp(iqr)\chi^{0}(q,p)\label{withoutden}\\
 & = & -\left[\begin{array}{c}
\int_{C_{2p}^{+}}dq\, q\exp(iqr)\chi^{0}(q,p)-\int_{C_{2p}^{-}}dq\, q\exp(iqr)\chi^{0}(q,p)\\
\int_{C_{-2p}^{+}}dq\, q\exp(iqr)\chi^{0}(q,p)-\int_{C_{-2p}^{-}}dq\, q\exp(iqr)\chi^{0}(q,p)dq\end{array}\right]\end{eqnarray}

Let us define, in analogy to Eq. (\ref{discontin}) \begin{eqnarray}
\chi^{0}(C_{-2p}^{+}) & \equiv & \lim_{\epsilon\rightarrow0}\chi^{0}(-2p-\epsilon+iq_{i},p)\nonumber \\
\chi^{0}(C_{-2p}^{-}) & \equiv & \lim_{\epsilon\rightarrow0}\chi^{0}(-2p+\epsilon+iq_{i},p)\end{eqnarray}

Due to Schwartz's principle, one has: \begin{equation}
\chi^{0}(q^{\ast},p)=\left[\chi^{0}(q,p)\right]^{\ast}=\chi^{0}(-q^{\ast},p)\end{equation}
 which implies that \begin{eqnarray}
\chi^{0}(C_{-2p}^{-}) & = & \left[\chi^{0}(C_{2p}^{+})\right]^{\ast}\nonumber \\
\chi^{0}(C_{-2p}^{+}) & = & \left[\chi^{0}(C_{2p}^{-})\right]^{\ast}\end{eqnarray}
 Substitution on (\ref{withoutden}) gives, after some algebra \begin{equation}
\int_{C}dq\, q\exp(iqr)\chi^{0}(q,p)=-2i\int_{0}^{\infty}dp\,\exp(-yr)Re\left[q\exp\left(i2pr\right)\Delta\chi^{0}(y,p)\right]\end{equation}
 Therefore \begin{eqnarray}
 &  & \int_{C}dq\, q\exp(iqr)\chi^{T}(q)=-2i\int_{0}^{\infty}dp\, g(p)\int_{0}^{\infty}dy\,\exp(-yr)\nonumber \\
 &  & Re\left[(2p+i)\exp\left(i2pr\right)\Delta\chi^{0}(y,p)\right]\end{eqnarray}

By repeating the above procedure, one can obtain the result:

\begin{eqnarray}
 &  & \int_{C}dq\, q\exp(iqr)\left[\chi^{T}(q)\right]^{2}=-4i\int_{0}^{\infty}dp\, g(p)\int_{0}^{\infty}dy\,\exp(-yr)\nonumber \\
 &  & Re\left[(2p+iy)\exp\left(i2pr\right)\Delta\chi^{0}(y,p)\chi^{T}(2p+iy)\right]\end{eqnarray}
 which can be easily generalized to the following formula \begin{eqnarray}
 &  & \int_{C}dq\, q\exp(iqr)\left[\chi^{T}(q)\right]^{n}=-2ni\int_{0}^{\infty}dp\, g(p)\int_{0}^{\infty}dy\,\exp(-yr)\nonumber \\
 &  & Re\left[(2p+iy)\exp\left(i2pr\right)\Delta\chi^{0}(y,p)\left[\chi^{T}(2p+iy)\right]^{n-1}\right]\label{intpowchi}\end{eqnarray}

Again, the necessary estimates on the above functions, in order to
justify all the steps, can be done in a direct albeit long procedure.

With the help of the previous equations, we can finally proceed with
Eq. (\ref{I(r)}).

\begin{equation}
I(r)=Im\int_{-\infty}^{+\infty}dq\,\frac{q\exp(iqr)}{\nu(q)}=Im\int_{-\infty}^{+\infty}dq\,\frac{q\exp(iqr)}{q^{2}-\lambda\chi^{T}(q)}\end{equation}
 To this end, we make an expansion of the denominator in powers of
$\lambda$. Using the result of Eq. (\ref{intpowchi}) we obtain,
after a straightforward calculation: \begin{equation}
I(r)=-2\int_{0}^{\infty}dp\, g(p)\int_{0}^{\infty}dy\,\exp(-yr)Re\left[q\exp\left(i2pr\right)\frac{\lambda\Delta\chi^{0}(y,p)}{\nu(q)^{2}}\right]\end{equation}
 Here \begin{equation}
q=2p+iy\end{equation}
 We can now use the above result to obtain an approximate expression
for $I(r)$ as $r\rightarrow\infty.$ In this case, due to the fast-decaying
exponential, is is enough to consider only small values of $y$. To
the leading order in $y$ we have, then: \begin{eqnarray}
\Delta\chi^{0}(y,p) & \simeq & -\frac{my}{4\pi}\nonumber \\
\nu(2p+iy) & \simeq & \nu(2p)\end{eqnarray}
 and one easily obtains \begin{equation}
I(r)\simeq\frac{\lambda m}{\pi r^{2}}\int_{0}^{\infty}dp\, g(p)\frac{p\cos(2pr)}{\nu(2p)^{2}}\label{asymptotic}\end{equation}
 This formula is valid, for an arbitrary temperature, at sufficiently
large distances and is specially suited, in contrast to the initial
expression Eq. (\ref{I(r)}), to low temperatures. Indeed, as $T\rightarrow0$
the function $g(p)$ is strongly peaked around the Fermi momentum
$p_{F}$ and allows for a fast convergence in the above expression.
Within this limit, therefore, we can make further approximations,
namely: \begin{equation}
I(r)\simeq\frac{\lambda m}{\pi r^{2}}\frac{p_{F}}{\nu(2p_{F})^{2}}\int_{0}^{\infty}dp\, g(p)\cos(2pr)\label{superpos}\end{equation}
 After some algebra, one arrives to the final result \begin{equation}
V(r)\simeq8\frac{e^{4}}{r^{2}}\frac{m^{2}T}{\nu(2p_{F})^{2}}\frac{\cos(2p_{F}r)}{\sinh\left(\frac{2\pi mTr}{p_{F}}\right)}\label{FriedT}\end{equation}
 In this way, we have obtained an oscillatory behavior for $V(r)$
at finite temperature. Whenever $\frac{2\pi mTr}{p_{F}}\gg1,$ one
can approximate $\sinh\left(\frac{2\pi mTr}{p_{F}}\right)\simeq\frac{1}{2}\exp\left(\frac{2\pi mTr}{p_{F}}\right),$
giving raise to an exponential suppression of Friedel oscillations
at finite temperature. This is a very well-known result, which can
be obtained using e.g. some modification of Lighthill's method to
finite temperature (see, for example \cite{DPS89,DGP94}). Such methods,
however, immediately raise the question of what the possible origin
of the oscillations is, as discussed in the introduction. Our method
is based on the integration of $\chi^{0}(q,p)$ on the complex plane,
by interchanging the limits of integration (see (\ref{interchange})).
In the next section we give an intuitive description of this procedure,
leading to (\ref{FriedT}).

\section{Discussion}

In this paper, we have revisited the method to obtain the screened
potential on an electron gas at finite temperature using the properties
of the zero-temperature polarizability. We analyzed with detail the
mathematical properties of this function on the complex plane. The
formula we obtained for the potential in the low-temperature regime
coincides with previous results in the literature, giving an oscillatory
function, which is damped as temperature increases. The existence
of oscillations at non-zero, but sufficiently low temperatures, shows
that the limit $T\rightarrow0$ is smoothly reached in the model,
and does not represent a pathology. However, the explanation based
on the Kohn singularity fails to account for the persistence of the
oscillations. In fact, the electron distribution is smeared out over
a range of the order $\sim T$ in energies, and therefore the Kohn
singularity disappears. On the other hand, for $T>0$ the function
$\chi^{T}(q)$ becomes continuous, so that the argument we used at
zero temperature does not apply to explain Friedel oscillations at
finite temperature.

In fact, as we have learned from the previous section, the function
$\chi^{T}(q)$ is non-analytical everywhere on the complex plane.
From an intuitive point of view, we can regard Eq. (\ref{chiT}) as
a superposition of a family (as $p$ varies) of functions $\chi^{0}(q,p),$
each one having discontinuities at $q=\pm2p$. This sum is spread
out, due to the weight function $g(p)$, over a range $\sim T$ in
energies. By interchanging the order of the integration, as in (\ref{interchange}),
we obtain a superposition of oscillatory terms, Eq. (\ref{superpos}),
which results in (\ref{FriedT}).

In our calculations, we have considered only a qualitative discussion
based on the RPA. It would be interesting to go beyond this approximation,
by introducing a local field correction. Unfortunately, to our knowledge,
there is no analytical formula which accounts for these effects in
the case of finite temperature. 
\begin{acknowledgments}
This work was supported by the Spanish Grants FPA2008-03373, FIS2010-16185
and 'Generalitat Valenciana' grant PROMETEO/2009/128.\end{acknowledgments}

\end{document}